\documentclass[]{JHEP}
\usepackage{epsfig}

\def\be{\begin{equation}}
\def\ee{\end{equation}}
\def\bea{\begin{array}}
\def\eea{\end{array}}
\def\beqa{\begin{eqnarray}}
\def\eeqa{\end{eqnarray}}
\def\beqas{\begin{eqnarray*}}
\def\eeqas{\end{eqnarray*}}

\def\bp{\begin{picture}}
\def\ep{\end{picture}}
\def\bc{\begin{center}}
\def\ec{\end{center}}
\def\bfig{\begin{figure}}
\def\efig{\end{figure}}

\def\bit{\begin{itemize}}
\def\eit{\end{itemize}}
\def\nn{\nonumber}
\def\f{\frac}
\def\sq{\sqrt}

\def\[{\left[}
\def\]{\right]}
\def\({\left(}
\def\){\right)}

\def\..{\left.}
\def\.{\right.}
\def\tl{\tilde}

\def\la{\leftarrow}

\def\tm{\times}

\def\da{\dagger}

\def\la{\lambda}

\def\al{\alpha}

\def\ep{\epsilon}

\def\de{\delta}
\def\De{\Delta}

\def\ga{\gamma}

\def\pr{\prime}
\def\eqv{\equiv}
\def\nm{\nonumber}

\title{Low-Scale $SU(4)_W$ Unification}
\author{Csaba Balazs$^1$,Tianjun Li$^{2,3}$, Fei Wang$^1$, Jin Min Yang$^2$ \\
 $^1$ School of Physics, Monash University, Melbourne Victoria 3800, Australia\\
 $^2$ Key Laboratory of Frontiers in Theoretical Physics,
      Institute of Theoretical Physics, Chinese Academy of Sciences,
      Beijing 100190, P. R. China \\
$^3$ George P. and Cynthia W. Mitchell Institute for Fundamental Physics,
     Texas A$\&$M University, College Station, TX 77843, USA
}

\abstract{ We embed the minimal left-right model $SU(2)_L\times
SU(2)_R\times U(1)_{B-L}$ into an $SU(4)_W$ gauge group, and break
the unified group via five-dimensional $S^1/(Z_2\tm Z_2)$
orbifolding.  Leptons are fitted into $SU(4)_W$ multiplets and
located on a symmetry preserving $O$ brane, while quarks are placed
onto an $O^\pr$ brane where the symmetry is broken. This approach
predicts $\sin^2\theta_W=0.25$ for the weak mixing angle at tree
level and leads to a rather low weakly ( strongly) coupled
unification scale of order $3 \tm 10^2$ TeV ( several TeV) with
supersymmetry, or as low as several TeV in the non-supersymmetric
case.  Another symmetry breaking chain with the low-energy gauge
group $SU(2)_L\tm U(1)_{3R} \tm U(1)_{B-L}$ can also give rise to a
weak mixing angle $\sin^2\theta_W=0.25$ at tree level after gauge
symmetry breaking by orbifolding.  Such theories with low-scale
unification have interesting phenomenological consequences.}

\begin{document}
\maketitle \indent
\newpage
\section{Introduction}

The standard model (SM) of electroweak interactions, based on the
spontaneously broken gauge symmetry $SU(2)_L{\times}U(1)_Y$, has
been extremely successful in describing phenomena below the weak
scale.  However, the SM leaves some theoretical and aesthetical
questions unanswered, two of which are the origin of parity
violation and the smallness of neutrino masses.  Both of these
questions can be addressed in the left-right model based on
$SU(2)_L\times SU(2)_R\tm U(1)_{B-L}$ \cite{mohapatra}. The
supersymmetric extension of this model \cite{susylr} is especially
intriguing since it automatically preserves R-parity.  This can lead
to a low energy theory without baryon number violating interactions
after R-parity is spontaneously broken.  However, in such left-right
models parity invariance and the equality of the $SU(2)_L$ and
$SU(2)_R$ gauge couplings is ad hoc and has to be put in by hand.
Only in grand unified theories, based on $SO(10)$ \cite{so10}, can
the equality of the two $SU(2)$ gauge couplings be naturally
guaranteed through gauge coupling unification.  But in these grand
unified theories the unification scale is usually much higher than
the weak scale. For example, in the supersymmetric $SU(5)$
\cite{su5} model the weak mixing angle is predicted to be $3/8$ at
tree level while the measured value is $0.23$ at the weak scale. The
difference can only be bridged via a long renormalization evolution,
which in turn requires a rather high unification scale at about
$10^{16}$ GeV. This high-scale unification has the unsatisfactory
feature that a large energy-desert lies between the weak scale and
the unification scale. Therefore, it is interesting to explore the
unification of the left-right symmetries at low energy scales.

Novel attempts for the unification of the left-right symmetries have
been proposed in the literature, such as the $SU(4)_{PS}\tm SU(4)_W$
or the $SU(4)_W\tm U(1)_{B-L}$ models \cite{shafi,nandi,lwy}.
However, in these new unification models the weak mixing angle
either can not be predicted (in $SU(4)_W\tm U(1)_{B-L}$ the weak
mixing angle is arbitrary) or predicted as $3/8$ at tree level, implying
a relatively high unification scale.  Besides, in order to
accommodate matter unification, mirror fermions are necessarily
introduced in order to fill each $SU(4)_W$ multiplet.  The
problems of these models are similar to the difficulty in the
$SU(3)_W\tm U(1)$ extension \cite{su3}, which uses $SU(3)_W$ to
unify the SM groups $SU(2)_L\tm U(1)_Y$.

With orbifold  gauge symmetry breaking (OGSB)
\cite{orbifold1, orbifold2, hebecker, hebecker2, Li:2001qs,
Li:2001wz}, we can achieve gauge interaction unification
while leaving matter fields partially unified or un-unified.  The
problem of the $SU(3)_W$ unification can be nicely tackled in this
approach \cite{su3l,su3h,su3k}.  In this work, we propose the use
of an $SU(4)_W$ group to unify the left-right gauge couplings on a
$S^1/(Z_2\tm Z_2)$ orbifold, in which leptons are fitted into
$SU(4)_W$ multiplets and located on the symmetry-preserving
$O$ brane while quarks are placed onto an $O^\pr$ brane with broken
symmetry.  This model predicts the weak mixing angle to be
$\sin^2\theta_W=0.25$ at tree level and achieves gauge coupling
unification at the order of $10^2$ TeV in supersymmetric cases and
several TeV in non-supersymmetric cases.

The content of this work is organized as follows. In Sec.
\ref{sec-2} we discuss $SU(4)_W$ left-right unification in the
supersymmetric (SUSY) context, focusing on gauge symmetry breaking
on the five-dimensional orbifold. In Sec. \ref{sec-3} we examine the
gauge coupling running and unification, especially the
compactification scale from the weak mixing angle. In Sec.
\ref{sec-4} we discuss another $SU(4)_W$ symmetry breaking chain
into $SU(2)_L\tm U(1)_{3R}\tm U(1)_{B-L}$. Sec. \ref{sec-5} contains
our conclusions.

\section{Brief Review of Orbifold Gauge Symmetry Breaking}

We consider a five-dimensional space-time ${\cal M}_4{\tm}
S^1/(Z_2{\tm}Z_2)$ comprising of Minkowski space ${\cal M}_4$ with
coordinates $x_{\mu}$ and the orbifold $S^1/(Z_2{\tm}Z_2)$ with
the coordinate $y \eqv x_5$. The orbifold $S^1/(Z_2{\tm}Z_2)$ is
obtained from $S^1$ by moduling the equivalent classes:
\beqa
P:~y{\sim} -y~,~~~~~~~~~P^{\pr}:~y^{\pr}\sim -y^{\pr}~,
\eeqa
with
$y^{\pr} \eqv y + \pi R/2$. There are two inequivalent 3-branes
located at $y=0$ and $y=\pi R/2$ which are denoted by $O$ and
$O^{\pr}$, respectively.

The five-dimensional $N=1$ supersymmetric gauge theory has 8 real supercharges,
corresponding to $N=2$ supersymmetry in four dimensions. The vector multiplet
physically contains a vector boson $A_M$ where $M=0, 1, 2, 3, 5$,
two Weyl gauginos $\lambda_{1,2}$, and a real scalar $\sigma$.
In the four-dimensional
$N=1$ language, it contains a vector multiplet $V(A_{\mu}, \lambda_1)$ and
a chiral multiplet $\Sigma((\sigma+iA_5)/\sqrt 2, \lambda_2)$ which
transform
in the adjoint representation of the gauge group.
The five-dimensional hypermultiplet has two physical complex scalars
$\phi$ and $\phi^c$, a Dirac fermion $\Psi$, and can be decomposed into
two 4-dimensional chiral multiplets $\Phi(\phi, \psi \equiv \Psi_R)$
and $\Phi^c(\phi^c, \psi^c \equiv \Psi_L)$, which transform as
each others conjugates under gauge transformations.

The general action \cite{nima2} for the gauge fields and their
couplings to the bulk hypermultiplet $\Phi$ is
\begin{eqnarray}
S&=&\int{d^5x}\frac{1}{k g^2}
{\rm Tr}\left[\frac{1}{4}\int{d^2\theta} \left(W^\alpha W_\alpha
+{\rm h.c.}\right) \right.\nonumber\\
&&~~~~~~~~~~~~~~~~
\left.+\int{d^4\theta}\left((\sqrt{2}\partial_5+ {\bar \Sigma })
e^{-V}(-\sqrt{2}\partial_5+\Sigma )e^V+
\partial_5 e^{-V}\partial_5 e^V\right)\right] \nonumber\\
&+& \int{d^5x} \left[ \int{d^4\theta} \left( {\Phi}^c e^V {\bar \Phi}^c
+{\bar \Phi} e^{-V} \Phi \right)
+ \int{d^2\theta} \left( {\Phi}^c (\partial_5 -{1\over {\sqrt 2}} \Sigma)
\Phi + {\rm h. c.} \right)\right]~~
\label{VD-Lagrangian}
\end{eqnarray}
where $Tr(T^aT^b)=k\delta^{ab}$.

 Because the action is invariant under the parity operation $P$,
under this operation, the vector multiplet transforms as
\begin{eqnarray}
V(x^{\mu},y)&\to  V(x^{\mu},-y) = P V(x^{\mu}, y) P^{-1}
~,~\,
\end{eqnarray}
\begin{eqnarray}
 \Sigma(x^{\mu},y) &\to\Sigma(x^{\mu},-y) = - P \Sigma(x^{\mu},
y) P^{-1}~.~\,
\end{eqnarray}
If the hypermultiplet belongs to the fundamental or anti-fundamental
representations, since $P=P^{-1}$, we have
\begin{eqnarray}
\Phi(x^{\mu},y)&\to \Phi(x^{\mu}, -y)  = \eta_{\Phi} P
\Phi(x^{\mu},y) ~,~\,
\end{eqnarray}
\begin{eqnarray}
\Phi^c(x^{\mu},y) &\to \Phi^c(x^{\mu}, -y)  = -\eta_{\Phi} P
\Phi^c(x^{\mu},y) ~.~\,
\end{eqnarray}
Alternatively, if the hypermultiplet belongs to the symmetric, anti-symmetric
or adjoint representations,  we have
\begin{eqnarray}
\Phi(x^{\mu},y)&\to \Phi(x^{\mu}, -y)  = \eta_{\Phi} P
\Phi(x^{\mu},y) P ~,~\,
\end{eqnarray}
\begin{eqnarray}
\Phi^c(x^{\mu},y) &\to \Phi^c(x^{\mu}, -y)  = -\eta_{\Phi} P
\Phi^c(x^{\mu},y) P ~,~\,
\end{eqnarray}
where $\eta_{\Phi} = \pm 1$.

Similar results hold for the parity operation $P'$, we just need to make the following
replacements in the above equations:
\begin{eqnarray}
P \longrightarrow P'~,~~~~~~~ \eta_{\Phi} \longrightarrow \eta'_{\Phi} ~.~\,
\end{eqnarray}

The gauge symmetry and supersymmetry can be broken by choosing suitable
representations for $P$ and $P'$.
For a field $\phi$, in
the representation of unbroken gauge symmetry, we obtain
the following transformation
\beqa
\phi(x_{\mu},y)& \to &  \phi(x_{\mu},-y) =  p_{\phi} \phi(x_{\mu},y) ~,\\
\phi(x_{\mu},y^{\pr})& \to & \phi(x_{\mu},-y^{\pr})= p_{\phi}^{\pr}
\phi(x_{\mu},y^{\pr}) ~,~\,
\eeqa
where $p_{\phi}=\pm1$ and $p_{\phi}^{\pr} = \pm1$.
Introducing the notation
$\phi_{p_{\phi} p_{\phi}^{\pr}}$, we obtain the Kaluza-Klein (KK) mode expansions
as of such $\phi$ fields as follows
\beqa
\phi_{++}(x_{\mu},y)&=&\sum\limits_{n=0}^{+\infty}
  \sq{\f{1}{2^{\delta_{n,0}}\pi R}}\phi_{++}^{(2n)}(x_\mu)\cos\f{2 n y}{R} \, ,\\
\phi_{+-}(x_{\mu},y)&=&\sum\limits_{n=0}^{+\infty}
 \sq{\f{1}{\pi R}} \phi_{+-}^{(2n+1)}(x_\mu)\cos\f{(2n+1) y}{R} \, ,\\
\phi_{-+}(x_{\mu},y)&=&\sum\limits_{n=0}^{+\infty}
 \sq{\f{1}{\pi R}} \phi_{-+}^{(2n+1)}(x_\mu)\sin\f{(2n+1)y}{R} \, ,\\
\phi_{--}(x_{\mu},y)&=&\sum\limits_{n=0}^{+\infty}
 \sq{\f{1}{\pi R}} \phi_{--}^{(2n+2)}(x_\mu)\sin\f{(2n+2)y}{R} \, .
\eeqa
Here $n$ is an integer and the fields $\phi_{++}^{(2n)}(x_\mu)$,
$\phi_{+-}^{(2n+1)}(x_\mu)$, $\phi_{-+}^{(2n+1)}(x_\mu)$ and
$\phi_{--}^{(2n+2)}(x_\mu)$ respectively acquire a mass of
$2n/R$, $(2n+1)/R$, $(2n+1)/R$ and $(2n+2)/R$ upon compactification.
Only $\phi_{++}(x_{\mu},y)$ possesses a four-dimensional massless zero mode.
It is easy to see that $\phi_{++}$ and $\phi_{+-}$ are non-vanishing
at $y=0$, while $\phi_{++}$ and $\phi_{-+}$ are non-vanishing at $y=\pi R/2$.

\section{$SU(4)_W$ Unification of $SU(2)_L\times SU(2)_R\times U(1)_{B-L}$}
\label{sec-2}

In the minimal left-right model based on
$SU(2)_L{\times}SU(2)_R{\times}U(1)_{B-L}$, the discrete symmetry
ensuring identical $SU(2)_L$ and $SU(2)_R$ gauge couplings is put in
by hand.  In this work we use $SU(4)_W$ to unify the left-right
symmetries and interpret the origin of the parity invariance as the
residual discrete symmetry from the symmetry breaking of the
unification group. Since we intend to truly unify the left-right
gauge groups, it is not possible to fill all the matter fields into
$SU(4)_W$ multiplets without introducing mirror fermions because of
the different $U(1)_{B-L}$ charge assignments for quarks and
leptons.  So we opt for the OGSB mechanism with partial unification
for matter fields.

Starting from the five-dimensional $SU(3)_C\times SU(4)_W$ gauge theory,
we can choose the following $Z_2$ matrix representations for
$P$ and $P'$ in the adjoint representation of  $SU(3)_C\times SU(4)_W$:
\beqa
P&=&{\rm diag}(+1,+1,+1)\otimes {\rm diag}(+1,+1,+1,+1) ~,\\
P^{\pr}&=&{\rm diag}(+1,+1,+1)\otimes {\rm diag}(+1,+1,-1,-1)~.
\eeqa
The gauge symmetry $SU(4)_{W}$ is broken by boundary
conditions to $SU(2)_L \times SU(2)_R \times U(1)_{X}$
on the boundary $O^{\pr}$ brane while is preserved in the bulk and on the
$O$ brane. Consequently, the parity assignments for $V$ and $\Sigma$ are
\beqa
V({\bf 15})&=&{\bf (3,1)_0^{(+,+)}\oplus(1,3)_0^{(+,+)}
  \oplus(2,\bar{2})_2^{(+,-)}\oplus({\bar 2},2)_{-2}^{(+,-)}
  \oplus(1,1)_0^{(+,+)}}~, \\ \nn \\
\Sigma({\bf 15})&=&{\bf (3,1)_0^{(-,-)}\oplus(1,3)_0^{(-,-)}
  \oplus(2,\bar{2})_2^{(-,+)}\oplus({\bar 2},2)_{-2}^{(-,+)}
  \oplus(1,1)_0^{(-,-)}}~.
\eeqa

We place the lepton sector on the $O$ brane while keep the quark
sector on the $O^\pr$ brane. This means that only the leptons
are filled into $SU(4)_W$ multiplets
\begin{eqnarray}
({\bf 4}):~~ L={\rm diag} \left(\nu_L~,~e_L~,~e_L^c~,~-\nu^c_L\right)~.
\end{eqnarray}
Here $\phi_L^c{\equiv}(\phi^c)_L$ and the minus sign conforms to
our choice of $Q^a=(\nu_L^c, e_L^c)$ in $SU(2)_R$ representations
${\bf \bar{2}}$ and being related to its conjugate by
$Q_a=(e_L^c,-\nu_L^c)$ through the fully antisymmetric tensor
$Q^a=\epsilon^{ab}Q_b$.
From the $SU(4)_W$ fundamental
representation and its proper normalization follows that the
$U(1)_{X}$ charge assignment of the fundamental representation can
be written as
\begin{eqnarray}
 Y_{X}={\rm diag} \left( -1,-1,~1,~1\right)~.
\end{eqnarray}
From this we can identify $U(1)_X$ as $U(1)_{B-L}$. The
normalization of the gauge group $U(1)_{B-L}$ reads
\begin{eqnarray}
T_{B-L}=\frac{\sqrt{2}}{2}\f{Y_{B-L}}{2} ~.~\,
\end{eqnarray}
From the normalization condition follows the relation between the
gauge couplings of $SU(2)_L\tm SU(2)_R \tm U(1)_{B-L}$ and $SU(4)_W$
\begin{eqnarray}
 g^2_{B-L}=\frac{1}{2}g_4^2~,~~~~~~~~g_L^2=g_R^2=g_4^2~,
\end{eqnarray}
which holds at the $SU(4)_{w}$ unification scale. Hence we can
predict the tree-level weak mixing angle as
\begin{eqnarray}
\sin^2\theta_W=\frac{g_{B-L}^2}{g_L^2+2g_{B-L}^2}=\frac{1}{4}~.
\end{eqnarray}

To induce Yukawa couplings for the $SU(4)_W$ multiplet leptons, we
can introduce bulk Higgs fields in the $SU(4)_W$ antisymmetric
representation $\Phi_{ab}({\bf \bar{6}})$ and
 symmetric representations $\Delta_{ab}^i({\bf {10}})$
and $\Delta_{ab}^i({\bf \overline{10}})$ \footnote{If we introduce only
$\Phi_{ab}({\bf \bar{6}})$ to give charged lepton masses, the
allowed Yukawa couplings have the form $y_{[ij]}L^{i}_{[a}
L^j_{b]}\Phi^{[ab]}$ with index $ab$( and $ij$)  being
antisymmetric. Such Yukawa couplings lead to $m_e=0$ and
$m_\mu=m_\tau$, which is unrealistic. Note that it is also possible
to introduce four
 Higgs hypermultiplets in the symmetric representation
$\Delta_{ab}^i({\bf {10}})$ and  $\Delta_{ab}^i({\bf \overline{10}})$
in this scenario.}.
In four dimensions there are eight $N=1$ chiral multiplets
$\Phi,(\Phi^{c})$, $\Delta^i$, $(\De^{c})^i$ ($i=1,~2,~3$).
 We assign the boundary conditions for the Higgs fields as
\beqa
\eta_{\Phi}=1,    ~~\eta^\pr_{\Phi}=-1,~~
\eta_{\Delta^1}=1,~~\eta^\pr_{\Delta^1}=-1,~~
\eta_{\Delta^i}=1,~~\eta^\pr_{\Delta^i}=1~(i=2,3)
\eeqa
The parities of the Higgs fields in terms of the $SU(2)_L\times
SU(2)_R\times U(1)_{B-L}$ representation are given by \beqa
\Phi({\bf {\bar 6}})&=&{\bf (1,1)_{2}^{(+-)}\oplus(1,1)_{-2}^{(+-)}
 \oplus(\bar{2},\bar{2})_{0}^{(++)}}~,~~~\nm \\
\Delta^1({\bf {\bar 10}})&=&{\bf
(\bar{3},1)_{-2}^{(+-)}\oplus(1,\bar{3})_{2}^{(+-)}
  \oplus(\bar{2},\bar{2})_0^{(++)}}~,~\nm\\
\Delta^2({\bf {\bar 10}})&=&{\bf
(\bar{3},1)_{-2}^{(++)}\oplus(1,\bar{3})_{2}^{(++)}
  \oplus(\bar{2},\bar{2})_0^{(+-)}}~,~\nm\\
\Delta^3({\bf { 10}})&=&{\bf
({3},1)_{2}^{(++)}\oplus(1,{3})_{-2}^{(++)}
  \oplus({2},{2})_0^{(+-)}}~.~\,
\eeqa
Under these parity assignments the conjugate chiral fields
$(\Phi^c),~(\Delta^c)^i$ ($i=1,~2,~3$) have no zero modes
and irrelevant to the low energy phenomenology. The zero
modes form two $SU(2)_L$ and $SU(2)_R$ triplets with opposite
$U(1)_{B-L}$ quantum numbers and two bi-doublets $(2,2)$ with
vanishing $U(1)_{B-L}$ quantum numbers which give exactly the Higgs
field contents of the supersymmetric left-right model.

As the leptons are fitted into the $SU(4)_W$ multiplets, we can
write down their Yukawa interactions with the bulk Higgs fields.
Since the leptons are placed on the $O$ brane, it is obvious that
they are invariant under $P$ transformation. The transformation
property for the leptons under $P^\pr$ is determined by the
requirement that the operators on the $O$ brane must transform
covariantly under $P^\pr$, otherwise the gauge symmetry preserved at
 $y=0$ will not be preserved at the $y=\pi R$ brane.
From the kinetic terms of the leptons we can
get the transformation property of the leptons under $P^\pr$ as
\beqa
 P^{\pr}:\(\nu_L~e_L~e_L^c~-\nu_L^c\)\to\pm(+,+,-,-) ~.~\,
\eeqa
The transformation of the Yukawa interactions under $P^\pr$ is
\beqa
 P^\pr:&& \sum\limits_{ij} Y_{ij}  L_{[i}^a L_{j]}^b \Phi_{[ab]}({\bf \bar{6}})=- ~,
\eeqa
where $(i,j)$ is antisymmetric, and
\beqa
P^\pr:&& \sum\limits_{ij} Y_{ij} L_i^a L_j^b \Delta_{ab}^{1} ({\bf
\overline{10}})=-~,~\\
P^\pr:&& \sum\limits_{ij} Y_{ij} L_i^a L_j^b \Delta_{ab}^{2} ({\bf
\overline{10}})=+~,~
\eeqa
where $i,j$ are the family indices. So we can write the Yukawa
interactions as
\beqa
{\cal L}_5&=&\int d^2\theta \sqrt{\pi R}\[
  \f{1}{2}\left\{\delta(y)-\delta(y-\pi R)\right\}\sum\limits_{ij}\(
  Y^1_{[ij]}  L_i^a L_j^b\Phi_{[ab]}^1+Y^2_{ij} L_i^a
  L_j^b\Delta_{ab}^1 \)\right. \nonumber \\
&&~~~~~~~~~~~~~~
\left. + \f{1}{2}\left\{\delta(y)+\delta(y-\pi
R)\right\}\sum\limits_{ij}Y^3_{ij}L_i^a L_j^b\Delta_{ab}^2 \] ~.~\,
\eeqa

After integrating out the fifth dimensional coordinate, we get the
Yukawa couplings in four dimensions
\beqa
{\cal L}_4&=&\sum\limits_{n=0}^{\infty}\int
  d^2\theta\sum\limits_{ij} \[\f{}{} \right. \f{1}{\sqrt{2^{n,0}}}
  \( \right. y^{1[ij]}
  L_{[i} L_{j]}^c\phi^{(2n)}_1+y^{2ij}L_i L_j^c\phi^{(2n)}_2 \nn \\
&&~~~~~~
 +y^{3ij}L_i L_j\Delta_1^{(2n)}
 +y^{4ij}(L_i^c) L_j^c\Delta_2^{(2n)}\left.\left. \) \f{}{}\]+h.c. ~,~\,
\eeqa
where the lepton $SU(4)_W$
multiplets are decomposed as ${\bf 4}=(L~L^c)$, the bi-doublet Higgs
fields $\phi_1$ and $\phi_2$ belong to the $(2,2)_0$ representations of
$SU(2)_L\tm SU(2)_R\tm U(1)_{B-L}$, and the triplets $\Delta_1$ and
$\Delta_2$ belong to $(3,1)_{-2}$ and $(1,3)_{2}$ representations.
The interactions for the zero modes of the Higgs fields are the
Yukawa couplings in the supersymmetric left-right model. Similarly,
we can write the couplings of the lepton multiplets with the vector
multiplet $V^a$ and the chiral multiplet $\Sigma^a$.

 Supersymmetry breaking can be realized via the Scherk-Schwarz mechanism
through the boundary conditions \cite{susyb,quiros,nomura}. It is
well known that $N=1$ supersymmetry in five dimensions possesses an
$SU(2)_R$ global R-symmetry under which the gauginos from the vector
multiplets $(\la_1,\la_2)$ and complex scalars $(\phi,\phi^{c\da})$
from hypermultiplets form $SU(2)_R$ doublets. The non-trivial twist
$T$ for translation with respect to $SU(2)_R$ R-symmetry can be
written as \cite{quiros,nomura}
 \beqa T=\exp\(-2\pi i \sigma_2\alpha\)~,~~~ \eeqa
with orbifolding projection \beqa P^\pr=\sigma_3~.~~~~~~~~ \eeqa
Besides, the symmetric Higgs bosons $\Delta^i({\bf 10})$(i=1,2,3)
have a $SU(3)$ flavor symmetry. Also, the consistent relation
between the translation and the orbifolding is
\beqa
\label{bconsistent}
 T{P^\pr}T=P^\pr~,~~~~ \eeqa
 where $P^\pr$ is the reflection according to $Z_2$(or $Z_2^\pr$),
 and $T$ is the translation
 \beqa
T \phi(x_\mu,y)=\phi(x_\mu,y+2\pi R)~.~~~~~
 \eeqa
We denote the translation operator $T$ corresponding to the global
$SU(3)$ flavor symmetry as follows \beqa T=\exp{\(2\pi i
\sum\limits_{a} T^a\theta^a\)}~,~~~~ \eeqa  where $T^a$ are $SU(3)$
generators. From formula (\ref{bconsistent}), we obtain
 \beqa
\left\{T^a\theta^a,P^\pr\right\}=0~.
 \eeqa
 For the following non-trivial $3\tm 3$ matrix
 \beqa
P^\pr=\left( \bea{cc}\pm 1&0\\0&\sigma_3\eea \right)~,~~~~
 \eeqa
  the most general form of $T$ can be described
by \beqa T=\exp\[{2\pi i (\gamma_0T^0+\ga_1T^1+\ga_2T^2})\] ~,\eeqa
with \beqa T^0&=&\left(\bea{ccc} ~0&~0&~0\\~0&~0&~1\\~0&~1&~0 \eea
\right)~,~~~~~~~~~~~~~~~~~~~T^1=\left(\bea{ccc}
~0&~0&~0\\~0&~0&-i\\~0&~i&~0
\eea\right)~,~\\
T^2&=&\left(\bea{ccc} ~0&~0&-i\\~0&~0&~0\\~i&~0&~0 \eea
\right)(P^\pr_{11}=1)~~{\rm or}~~~T^2=\left(\bea{ccc}
~0&-i&~0\\~i&~0&~0\\~0&~0&~0 \eea \right)(P^\pr_{11}=-1)~. \eeqa The
parameter $\gamma_0$ can be rotated away by the residue global
symmetry, so the twist for flavor $SU(3)$ compatible with
non-trivial $P^\pr$ can be written as \beqa T=\exp\[2\pi i
(\gamma_1T^1+\ga_2T^2)\]~. \eeqa
 In case of the $SU(3)$ flavor symmetry
for $\De^1({\bf \bar{10}}),\De^2({\bf \bar{10}}),\De^3({\bf 10})$,
the relative parity assignments under $P$ and $P^\pr$ are
 nontrivial with
\beqa P=\left( \bea{cc}1&0\\0&\sigma_3\eea \right)~,~
      P^\pr=\left(
\bea{cc}-1&0\\0&\sigma_3\eea \right)~.~~~~ \eeqa So the twist
boundary condition $T$ compatible with both are \beqa T=\exp\(2\pi i
\gamma_1T^1\)~. \eeqa

 The most general
boundary conditions for the fields are
 \beqa
A^M(x^\mu,y+2\pi R)&=& A^M(x^\mu,y) ~,~ \\
\sigma(x^\mu,y+2\pi R)&=&  \sigma(x^\mu,y) ~,~ \\
\left(\bea{c} \la_1  \\
\la_2\eea \right)(x^\mu,y+2\pi R)&=&e^{-2\pi i\al \sigma_2}\left(\bea{c} \la_1 ~~ \\
\la_2\eea \right)(x^\mu,y) ~,~ \\
\left(\bea{c} \phi~  \\
\phi^{c\da}\eea \right)(x^\mu,y+2\pi R)&=&e^{-2\pi i\al \sigma_2}\left(\bea{c} \phi ~ \\
\phi^{c\da}\eea \right)(x^\mu,y) ~,~\\
\left(\bea{ccc} \tilde{\de}_1~&\tilde{\de}_2~&\tl{\de}_3~
 \\
\tl{\de}^{c\da}_1~&\tl{\de}_2^{c\da}~&\tl{\de}^{c\da}_3\eea
\right)(x^\mu,y+2\pi R)&=&\left(\bea{ccc}
\tl{\de}_1~&\tl{\de}_2~&\tl{\de}_3 ~ \\
\tl{\de}^{c\da}_1~&\tl{\de}_2^{c\da}~&\tl{\de}^{c\da}_3
\eea \right)e^{2\pi i \ga_1T^1}(x^\mu,y) ~,~ \\
\left(\bea{ccc} \de_1~&\de_2~&\de_3 ~ \\
{\de}^{c\da}_1~&{\de}_2^{c\da}~&{\de}^{c\da}_3\eea
\right)(x^\mu,y+2\pi R)&=&e^{-2\pi i\al \sigma_2}\left(\bea{ccc}
\de_1~&\de_2~&\de_3 ~ \\
{\de}^{c\da}_1~&{\de}_2^{c\da}~&{\de}^{c\da}_3
\eea \right)e^{2\pi i \ga_1T^1 }(x^\mu,y) ~.~ \,
 \eeqa
 Here we denote the components of chiral supermultiplets $\De^i({\bf 10})$ as
  $(\de^i({\bf 10}),\tilde{\de}^i({\bf 10}))$ with their conjugate chiral supermultiplets
$\De^{ic}({\bf \bar{10}})$ as
  $({\de}^{ic}({\bf 10}),\tilde{\de}^{ic}({\bf 10}))$. The complex scalar components
  for hypermultiplets  ($\Phi({\bf 6}),\Phi^c({\bf \bar{6}})$) are denoted as
  $(\phi,\phi^{c\da})$.

 We now consider the modes expansion of the fields with
respect to the previous Scherk-Schwarz type boundary conditions. For
simplicity, we write explicitly only the relative $P$ and $P^\pr$ parity
assignments under orbifolding projections
\beqa
&&\left(\bea{c} \la_1^{(++)}\\
\la_2^{(--)}\eea \right)(x^\mu,y)=\nn  \\
&&~~~~~~~~~~~~~~~~~~ \sum\limits_{n=0}^\infty  e^{-i\al \sigma_2 y/R } \left(\bea{c}  \sq{\f{1}{2^{\delta_{n,0}}\pi R}}(\la_{1}^{(++)})^{(2n)}(x_\mu)\cos\f{2 n y}{R}\\
~~ \sq{\f{1}{\pi R}} (\la_{2}^{(--)})^{(2n+2)}(x_\mu)\sin\f{(2n+2)y}{R}\eea \right) ~,~ \\
&&\left(\bea{c} \phi^{(++)}\\
\(\phi^{c\da}\)^{(--)}\eea \right)(x^\mu,y)=\nn  \\
&&~~~~~~~~~~~~~~~~~~~\sum\limits_{n=0}^\infty  e^{-i\al \sigma_2 y/R } \left(\bea{c}  \sq{\f{1}{2^{\delta_{n,0}}\pi R}}\phi_{1++}^{(2n)}(x_\mu)\cos\f{2 n y}{R} \\
 \sq{\f{1}{\pi R}} \phi_{2--}^{(2n+2)}(x_\mu)\sin\f{(2n+2)y}{R}\eea \right) ~,~ \\
&&\left(\bea{ccc} \tilde{\de}_1^{+-}~&\tilde{\de}_2^{++}~&\tl{\de}_3^{--}  \\
\tl{\de}^{c\da-+}_1~&\tl{\de}_2^{c\da--}~&\tl{\de}^{c\da++}_3\eea
\right)(x^\mu,y)=\nn  \\
 &&~~~~~~~~~~~~~~\sum\limits_{n=0}^\infty\left(\bea{ccc}
(\tl{\de}_1)^{+-}_{(2n)}~&(\tl{\de}_2)^{++}_{(2n)}~&(\tl{\de}_3)^{--}_{(2n+2)} \\
(\tl{\de}^{c\da}_1)^{-+}_{(2n+2)}~&(\tl{\de}_2^{c\da})^{--}_{(2n+2)}~&(\tl{\de}^{c\da}_3)^{++}_{(2n)}
\eea \right)e^{i \ga_1T^1y/R}(x^\mu,y) ~,~ \\
&&\left(\bea{ccc}{\de}_1^{+-}~&{\de}_2^{++}~&{\de}_3^{--}\\
{\de}^{c\da-+}_1~&{\de}_2^{c\da--}~&{\de}^{c\da++}_3\eea
\right)(x^\mu,y)=\nn  \\
&&~~~~~~~~~~~~~\sum\limits_{n=0}^\infty e^{-i\al \sigma_2 y/R
}\left(\bea{ccc}
({\de}_1)^{+-}_{(2n)}~&({\de}_2)^{++}_{(2n)}~&({\de}_3)^{--}_{(2n+2)}\\({\de}^{c\da}_1)^{-+}_{(2n+2)}~&({\de}_2^{c\da})^{--}_{(2n+2)}~&({\de}^{c\da}_3)^{++}_{(2n)}
\eea \right)e^{i
\ga_1T^1y/R}(x^\mu,y)~,~\nn\\
 \eeqa \normalsize
 in which we represent the symmetric $\De^i({\bf \bar{10}})$(i=1,2,3) by its
 $({\bf \bar{3},1})_{-2} $ modes.
 The zero modes from the orbifold projection can get mass terms from the previous Scherk-Schwarz boundary
conditions
 \beqa
{\cal L}=
&-&\f{\al}{2R}\sum\limits_{a}\(\la_0^a\la_0^a+h.c.\)-\f{\al}{R}\(\tl{\Delta}_L^1\tl{\Delta}_L^2+\tl{\Delta}_R^1\tl{\Delta}_R^2+h.c.\)
-\f{\al^2}{R}\(Tr\(\Phi_1^\da\Phi_1\)+Tr\(\Phi_2^\da\Phi_2\)\)\nn\\
&-&\(\f{\al^2}{R^2}+\f{\ga^2}{R^2}\)\(Tr\(\Delta_L^{1\da}\Delta_L^1\)+Tr\(\Delta_L^{2\da}\Delta_L^2\)+Tr\(\Delta_R^{1\da}\Delta_R^1\)
+Tr\(\Delta_R^{2\da}\Delta_R^2\)\)\nn\\
&+&\f{2\al\ga}{R^2}\({\Delta}_L^1{\Delta}_L^2+{\Delta}_R^1{\Delta}_R^2+h.c.\)~.
\eeqa
 Here triplets $\Delta_L^1\[({\bf \bar{3},1})_{(-2)}\],\Delta_R^1\[({\bf
 1,\bar{3}})_{(2)}\]$ are zero modes from $\De^2({\bf \bar{10}})$
while triplets $\Delta_L^2\[({\bf {3},1})_{(2)}\],\Delta_R^2\[({\bf
 1,{3}})_{(-2)}\]$ from $\De^3({\bf {10}})$.
The bi-doublets $\Phi_1({\bf \bar{2},\bar{2}})_0$
are zero modes from $\Phi({\bf \bar{6}})$ while
 $\Phi_2({\bf \bar{2},\bar{2}})_0$ from $\De^1({\bf \bar{10}})$.
The gauge index $a$ runs over the left-right gauge group
 $SU(3)_c$, $SU(2)_L$, $SU(2)_R$, and $U(1)_{B-L}$.
The continuous parameters $\al$ and
$\ga$ can be chosen to be $\al,\ga \ll 1$ \cite{nomura} or
 $\al,\ga\sim {\cal O}(1)$ \cite{quiros}. We chose the former case with $\al,\ga \ll1$.
  If the scale of the supersymmetry breaking soft  mass terms $\al/R$
and $\ga/R$ is chosen to be at the order of the electroweak scale,
we can get the relation $M_S<M_R$. Otherwise if the scale for the
supersymmetry breaking soft mass terms is higher than $M_R$ which
will not explain the gauge hierarchy problem, $M_R < M_S$ is also
possible.
 In our case the matter contents are placed at
the orbifold fix point so that no tree-level mass terms are
generated through orbifolding. However, the sfermions masses can be
radiatively generated  through renormalization group equations below
the compactification scale. Since such interactions are almost
flavor universal, the supersymmetric flavor problems can be
solved.There is a lot of freedom to tune the complicate Higgs
potential to break the left-right symmetry down to $U(1)_Q$
directly. In supersymmetric left-right models, sneutrinos can couple
to the Higgs sector which leads to spontaneously broken R-parity if
such sneutrino doublets acquire vacuum expectation values. The
couplings between the triplets and sneutrino which arise from the
Yukawa superpotential are rather arbitrary. Detailed discussion on
Higgs potential coupled to sneutrino doublets can be found in
Ref.~\cite{susylr,ji} which will not be discussed here. In SUSY
left-right cases, R-parity is automatically conserved.

The chiral anomaly cancellation in OGSB cases has been studied in
\cite{nima,sssz,pr,lee}. In our case with $S^1/(Z_2\tm Z_2)$ OGSB,
if the gauge anomaly in four dimensions is cancelled, the
five-dimensional fix-point gauge anomaly can be cancelled by
introducing appropriate bulk Chern-Simons terms with jumping
coefficients. At the fix point $O$, the gauge anomaly from the
lepton ${\bf 4}$ representation and the bulk Higgsinos in
representation ${\bf \bar{6},\overline{10},10}$ are cancelled by
the five-dimensional
 Chern-Simons terms. Such Chern-Simons terms also
cancel the quark contribution on the $O^\pr$ brane. At the fix point
$O^\pr$ we can see that the four-dimensional anomaly associated with
the bulk Higgsinos is cancelled automatically although the bulk fermion
contributions to the anomaly associated with the unbroken gauge
group add up.
\vspace*{0.5cm}

\noindent {\bf Alternative Models:~~}
 It is also possible to put the leptons into the bulk by
introducing mirror leptons and placing quarks on the broken symmetry
$O^\pr$ brane. We can introduce bulk hypermultiplets
$(F_L,F_R)$ in the $\mathbf{(1,4)}$ representation and
$(F_L^c,F_R^c)$ in the $\mathbf{(1,{\bar 4})}$ representation.
These multiplets are filled as:
\beqa
F_L&=&(L_L,X_L)~,~~~~~F_R=(X_L^c,L_L^c)~,~\\
F_L^c&=&(\bar{L}_L,\bar{X}_L)~,~~~~~F_R^c=(\bar{X}_L^c,\bar{L}_L^c)
~,~\, \eeqa where
$X_L^c,\bar{L}_L,\bar{X}_L^c$($X_L,\bar{L}_L^c,\bar{X}_L$)are left
(right) handed mirror leptons. $L_L$ and $L_L^c$ are left and right
handed leptons in minimal left right model, respectively. Lepton
doublets in the minimal left-right model can be obtained by
introducing the following parity assignments:
\begin{eqnarray}
\eta_{F_{L}}=1~,~~~\eta'_{F_{L}}=1~,~~~\eta_{F_R}=+1
~,~~~\eta'_{F_R}=-1~.~\,
\end{eqnarray}
Lepton $SU(2)_L$ doublets survive projections from
$F_L$ while lepton $SU(2)_R$ doublets from $F_R$. We can see from
the charge assignments of the bulk hypermultiplets that the tree
level weak mixing angle $\sin^2\theta_W=0.25$ still holds in this
scenario.

The Higgs sector can be placed in the bulk or localized on the broken
symmetry $O^\pr$ brane. In the latter case,
we need two bi-doublets $({\bf 2,2,0})$, two $SU(2)_L$
triplets $({\bf 3,1,\pm 2})$ and two $SU(2)_R$ triplets $({\bf
1,3,\pm 2})$ in left-right gauge group $SU(2)_L\tm SU(2)_R \tm
U(1)_{B-L}$ representations. The case of the bulk Higgs is almost
identical to that of the previous case, so we do not discuss it in detail.

It is also possible to put quarks in the bulk while locate leptons on
the $O^\pr$ brane. Then we introduce mirror quarks $\bar{Q},\bar{Q}^c$ to
fill $SU(3)_c\tm SU(4)_W$ representations as:
\beqa
{\bf (3,4)}={\rm
 diag}(U_L~,~\bar{Q}_L^c)~,~~~~~~~~~~~
{\bf (\bar{3},{4})}={\rm diag}(\bar{Q}~,~U_L^c)  ~,~\,
\eeqa
where $U_L^c=(d_L^c,-u_L^c)$ denote the ${\bf 2}$ representations in
 $SU(2)_R$. In this case the $U(1)_{B-L}$ charge for
 $SU(4)_W$ fundamental representation reads
 \beqa
Y_{B-L}={\rm diag}(~\f{1}{3},~\f{1}{3},-\f{1}{3},-\f{1}{3}) ~,~\,
 \eeqa
 which is normalized with respect to the $SU(4)_W$ generator $T_{B-L}$ as:
 \beqa
T_{B-L}=\f{3\sqrt{2}}{2}\f{Y_{B-L}}{2} ~.~\,
 \eeqa
 From the gauge coupling relations
 \beqa
g_{B-L}=\f{3\sqrt{2}}{2}g_4~,~~~~~~~~~~~~~~g_L=g_R=g_4 ~,~\,
\eeqa
we can get the tree level weak mixing angle
\beqa
\sin^2\theta_W=\f{g_{B-L}^2}{g_L^2+2g_{B-L}^2}=0.45 ~,~\,
\eeqa
which is not acceptable as a low-energy unification model.

\section{Gauge Coupling Running and Unification Scale}
\label{sec-3}

In this section we discuss the renormalization group equation
(RGE) running of the gauge couplings in the orbifold breaking case.
We consider only the simplest scenario without mirror fermions.
At the weak scale our inputs are \cite{PDG}
 \beqa
  M_Z&=&91.1876\pm0.0021 ~,~\,\\
  \sin^2\theta_W(M_Z)&=&0.2312\pm 0.0002 ~,~\,\\
  \alpha^{-1}_{em}(M_Z)&=&127.906\pm 0.019 ~,~\,\\
  \alpha_3(M_z)&=&0.1187\pm 0.0020 \, ,
 \eeqa
which fix the numerical values of the standard $U(1)_Y$ and
$SU(2)_L$ couplings at the weak scale
\beqa
\alpha_1(M_Z)&=&\f{\alpha_{em}(M_Z)}{\cos^2\theta_W}=(98.3341)^{-1}~,\\
\alpha_2(M_Z)&=&\f{\alpha_{em}(M_Z)}{\sin^2\theta_W}=(29.5718)^{-1}~.
\eeqa

 The RGE running of the gauge couplings reads
\beqa \f{d~\alpha_i}{d\ln E}=\f{b_i}{2\pi}\alpha_i^2~ \, , \eeqa
where $E$ is the energy scale and $b_i$ are the beta functions. At
the scale of the $SU(2)_R$ gauge boson mass $M_R$, the left-right
$SU(3)_C\times SU(2)_L\times SU(2)_R\times U(1)_{B-L}$ symmetry
breaks to the SM gauge group. From the symmetry breaking chain and
the normalization of the gauge field $(g_{B-L}Y_{B-L}/2)
A_{\mu}^{B-L}$ in the kinetic term, we obtain the relation
\beqa
\f{1}{e^2}=\f{1}{g_{2L}^2}+\f{1}{g_{2R}^2}+\f{1}{g_{B-L}^2} \, ,
\eeqa
from which we can calculate the coupling $g_{B-L}$ at the
scale $M_R$.

Note that in non-supersymmetric left-right models neutrino masses
arise by a Type I or Type II see-saw mechanism. In this case an
${\cal O}$(TeV) mass is unnatural for the $W_R$ gauge boson due to
the mixing term ${\rm
Tr}(\Phi\Delta_L\Phi^\dagger\Delta_R^\dagger)$. In the
supersymmetric left-right model such a mixing term is not allowed by
supersymmetry and thus a TeV-scale $W_R$ mass is realistic
\cite{ji}. We know that we need two bi-doublets to give tree-level
Cabibbo-Kobayashi-Maskawa (CKM) mixings in supersymmetric left-right
models. Thus, in the low energy limit, the electroweak symmetry
breaking Higgs sector is non-minimal, containing two
bi-doublets \footnote{Due to the left-right symmetry, the
left-handed triplets have the same masses as the right-handed
triplets which are at the order of the $M_R$ scale. On the other
hand, the vacuum expectation values (VEVs) for the left-handed
triplets are small because such VEVs will break $SU(2)_L$.}.  The
corresponding supersymmetric extension (below $M_R$) also contains
two bi-doublets.

Assuming that the $SU(2)_R$ gauge boson mass $M_R$ is in the range
$1 {\rm ~TeV}<M_R<M_C$ (where $M_C$ is the compactification scale)
and the mass of its superpartner falls in the range $200 {\rm
~GeV}<M_S<M_C$, we have two possibilities:
\begin{itemize}
\item[(i)] One possibility is that $M_S<M_R$. In this case $\al,\ga
\ll1$ with $\al/R,\ga/R$ at the order of the electro-weak scale.
Then the beta functions for the gauge couplings of
$U(1)_Y,SU(2)_L,SU(3)_c$ are given by \beqa
&&(b_1,b_2,b_3)=\(~\f{22}{3},-\f{8}{3},-7\) ~~~ {\rm for}~ M_Z<E<M_S~,\\
&&(b_1,b_2,b_3)=\(~12,~2,-3 \) ~~~~~~~ {\rm for}~ M_S<E<M_R~,
\eeqa
while for $\sqrt{2}U(1)_{B-L},SU(2)_L,SU(2)_R$, and $SU(3)_c$ they
are given by
\beqa
(b_1,b_2^L,b_2^R,b_3)=\(~8,~6,~6,-3 \) ~~~~~~~{\rm for}~ M_R<E<M_C~.
\eeqa
\item[(ii)] The other possibility is $M_R<M_S$.
In this case $\al/R,\ga/R$ are of order ${\cal O}(10)$ TeV  and the
gauge hierarchy problem is not solved by the high energy
supersymmetry. Then we have \beqa
(b_1,b_2,b_3)&=&\(~\f{22}{3},-\f{8}{3},-7\) ~~~~~~~~ {\rm for}~M_Z<E<M_R~, \\
(b_1,b_2^L,b_2^R,b_3)&=&\(~\f{10}{3},-\f{4}{3},-\f{4}{3},-7 \) ~~
{\rm for}~M_R<E<M_S ~, \\
(b_1,b_2^L,b_2^R,b_3)&=&\(8,6,6,-3 \) ~~~~~~~~~~~~~ {\rm for}~M_S<E<M_C ~.
\eeqa
\end{itemize}
Above the SUSY left-right scale the RGE running of the gauge couplings
receives contributions from KK modes
\beqa
 \alpha_i^{-1}(E)&=&\alpha_i^{-1}(M_R)+\f{b_i}{2 \pi}\ln\(\f{M_R}{E}\) +
 \f{b_{i,e}}{2 \pi}\sum\limits_{n=1}^{k}\ln\(\f{2 n }{E R}\)\Theta(E-\f{2n}{R})
  \nonumber \\ &&
+\f{b_{i,o}}{2 \pi}\sum\limits_{n=0}^{k}\ln\(\f{2 n+1 }{E R}\)\Theta(E-\f{2n+1}{R}) .
\eeqa
Here $\Theta(x)$ is the step function defined as
$\Theta(x)=1$ for $x\ge 0$ and $\Theta(x)=0$ for $x<0$. The beta
functions corresponding to the even and odd KK modes at 1-loop are
\beqa
 (b_{B-L,e}, b_{B-L,o})&=&(12,~0) ~, \\
 (b_{2,e}^L, b_{2,o}^L)&=&(~8,~4) ~, \\
 (b_{2,e}^R, b_{2,o}^R)&=&(~8,~4) ~,\\
 (b_{3,e}^R, b_{3,o}^R)&=&(-6,~0) ~,
\eeqa
that is, the RGE running of the $SU(2)_L$ and $SU(2)_R$ gauge
couplings are identical.

The existence of the symmetry breaking $O^\pr$ brane allows the
localized kinetic terms for the unbroken gauge group
$SU(2)_L{\times}SU(2)_R{\tm}U(1)_{B-L}$, which spoil the $SU(4)_W$
unification. The most general form of the gauge kinetic term is
given by
\small
\beqa
S=\int d^4x dy\left(\f{1}{4g_5^2}F_{MN}F^{MN}
 +\delta(y)\f{1}{4\bar{g}^2}F_{\mu\nu}F^{\mu\nu}+\delta(y-\f{\pi R}{2})
  \sum\limits_{i=0}^2\f{1}{4\tilde{g}_i^2}F_{\mu\nu}F^{\mu\nu}\right)~~~~
\eeqa
\normalsize
After integrating out the higher modes, the
gauge couplings of the zero modes are
\beqa
 \f{1}{g_i^2}=\f{\pi R}{2g_5^2}+\f{1}{\bar{g}^2}+\f{1}{\tilde{g}_i^2}~,
\eeqa where $g_0=\sqrt{2}g_{B-L}$, while $g_1,g_2$ correspond to
$g_{2L},g_{2R}$ coupling respectively. The term $\bar{g}^2$ is
irrelevant because it preserves $SU(4)_W$ unification and will not
affect the tree-level weak mixing angle. We can assume that the bulk
and brane kinetic terms have comparable strength \cite{su3h} at a
cut-off scale $\Lambda$ (higher than or equal to the unification
scale $M_U$). Since $g_5^2$ has mass dimension, we can estimate its
strength to be $g_5^2\Lambda$ at the cut-off scale, which implies
$g_5^2\Lambda\sim \tilde{g_i}^2$. We can see that at tree level the
$SU(4)_W$ violating term is suppressed by $M_C/\Lambda$ and hence
the effects can be neglected if $M_C \ll \Lambda$. Besides, it is
natural to set such localized gauge kinetic terms to zero at tree
level in a fundamental theory. Then for a weakly coupled theory,
such localized kinetic terms can only arise at loop level and thus
highly suppressed.

The one-loop corrections to the weak mixing angle come from the
$SU(4)_W$ violating effects but not from the $SU(4)_W$ conserving
effects. For the energy scale in the range $2N M_C < E < (2N+1) M_C$
with $N \gg 1$, we can estimate the RGE running by summing over the
contribution of the KK modes. Using Stirling's approximation
\beqa\label{stirling}
 \ln \(N!\)&\simeq& N\ln N-N+\f{1}{2}\ln \(2\pi N\) ~,
\eeqa
and
\beqa
 \ln\[1\times3\times...\times(2N-1)\]&=&\ln\[(2N)!\]-N\ln 2- \ln (N!)\nonumber\\
  &\simeq& \(N+\f{1}{2}\)\ln 2+ N\ln N-N~,
\eeqa
we can write
\small
\beqa
\label{appo}
 \alpha_i^{-1}(E) \simeq
 \alpha_i^{-1}(M_R)+\f{b_i}{2\pi}\ln\(\f{M_R}{E}\)
 -\f{1}{4\pi}(b_{i,o}+b_{i,e})\[\f{E}{M_C}-\ln 2\]
+\f{b_{i,e}}{4\pi}\ln\(\f{\pi E}{2 M_C}\)~~~~
\eeqa
\normalsize
Thus, after the
KK modes contributions are included, the RGE running of the gauge
couplings are proportional to $N=E/(2M_C)$, which is a power law
running (this agrees with the results of \cite{tony}). The relative
running of $SU(2)_L$ (identical to $SU(2)_R$) and $U(1)_{B-L}$ is
not affected by the $SU(4)_W$ conserving power-law running, instead
this running is logarithmic due to $SU(4)_W$ violating effects. In
OGSB cases, it is general to have
\beqa
 b_{B-L,e} + b_{B-L,o} = b_{L,e} + b_{L,o}  = b_{R,e} + b_{R,o}~,
\eeqa
which also holds in our case.  Due to the universally occurring
$b_o+b_e$ term, we can replace $b_e$ with $-b_o$ in the relative
running between the gauge couplings. The running of the minimal
left-right gauge couplings is given by
\beqa
\f{1}{g_i^2}(M_R)&\simeq&\f{1}{g_*^2}(M_U)+\f{a}{16\pi^2}\[\(\f{M_U}{M_C}\)-\ln 2\]
 +\f{\tl{b}_i}{8\pi^2}\ln\f{M_U}{M_C^\prime}+\f{\tl{c}_i}{8\pi^2}\ln\f{M_C^\pr}{M_R}~~~~
\eeqa
where $M_C^\pr=2M_C/\pi$, the coefficient $a$ which is
universal and $\tl{b}_i$ are given in our case by
\beqa
 a=b_{i,o}+b_{i,e} ~, ~~~~~~ \tl{b}_i=b_i-\f{1}{2}b_{i,e}~,~~~~~~\tl{c}_i=b_i~.
\eeqa
Then the Weinberg angle for non-SUSY cases is
\small
\beqa
 \sin^2\theta_W(M_Z)=
 \f{1}{4}
 -{\alpha_{em}}(M_Z)\[\f{\tl{b}_1-\tl{b}_2}{4\pi}\ln\f{M_U}{M_C^\prime}
   +\f{\tl{c}_1-\tl{c}_2}{4\pi}\ln\f{M_C^\pr}{M_R}
+\f{d_1-3d_2}{8\pi}\ln\f{M_R}{M_Z}\]~~~~
\eeqa
\normalsize
where $(d_1,d_2)$ are the one-loop beta functions for
$U(1)_Y,SU(2)_L$ in the energy range between $M_R$ and $M_Z$. The
Weinberg angle for SUSY cases is given by
\beqa
\sin^2\theta_W(M_Z)=\f{1}{4}-{\alpha_{em}}(M_Z)&&\[\f{\tl{b}_1-\tl{b}_2}{4\pi}\ln\f{M_U}{M_C^\prime}
 +\f{\tl{c}_1-\tl{c}_2}{4\pi}\ln\f{M_C^\pr}{M_R}
\right. \nn \\
&&\left. +\f{{d}_1-3d_2}{8\pi}\ln\f{M_R}{M_S}
   +\f{e_1-3e_2}{8\pi}\ln\f{M_S}{M_Z}\] ~,
\eeqa
where $(d_1,d_2)$ and $(e_1,e_2)$ denote the one-loop beta
functions for $U(1)_Y$ and $SU(2)_L$ for $M_R>E>M_S$ and $M_S>E>M_Z$,
respectively.

From the above formulas, we can estimate the unification scale once
the compactification scale $M_C^\pr$ is specified. In fact, we can
obtain the unification scale more precisely by taking into account
each KK-mode contribution step by step. In SUSY left-right
unification cases, we can see from the beta functions that
$\tilde{b}_i=2$ is universal for $SU(2)_L,SU(2)_R$ and $U(1)_{B-L}$.
It means that there is no relative running between the three gauge
couplings. However, Eq.~(\ref{appo}) is not valid if the unification
scale $M_U\sim N M_C$ satisfies $N\sim {\cal O}(1)$ with which the
summation approximation Eq.~(\ref{stirling}) is not valid. Thus, we
anticipate the unification occurs at the order of the
compactification scale if we require that the gauge coupling at
unification scale be not strong coupled (weakly coupled unification
). We can also identify the unification scale $M_U$ as the cut off
scale $\Lambda$ if the gauge coupling would be strongly coupled at
the unification scale. We know that $M_S$ is fixed to be within
several hundreds GeV in order to give an explanation of the gauge
hierarchy problem by supersymmetry (We will not discuss the
non-interesting case of high scale supersymmetry with $M_S>M_R$
here). The detailed numerical calculations show that there is a
fairly large parameter space for the values of $M_R$ and $M_C$ in
which the weakly coupled unification is possible. We find that the
compactification scale $M_C$ is required to be larger than $150$ TeV
in order to get successful weakly coupled gauge coupling
unification. While the larger the $M_C$, the lower the possible
value of $M_R$ that is allowed by the weakly coupled gauge coupling
unification. For example, the parameter $M_R$ is required to be
larger than $70$ TeV with $M_C=150$ TeV. While if $M_C=200$ TeV, the
allowed $M_R$ can be as low as 40 TeV. Fixing the left-right scale,
which is identified as the $SU(2)_R$ gauge boson masses, to
$M_R=100$ TeV, the sfermion mass $M_S=600$ GeV, and the
compactification scale $M_C=200$ TeV, we obtain \beqa
  \alpha_{B-L}^{-1}(M_R)=28.810705 ~,~ ~~~
  \alpha_{L}^{-1}(M_R)=\alpha_{R}^{-1}(M_R)=28.742994 ~,~
\eeqa and a weakly coupled unification scale \beqa M_U =323.5 {\rm
~TeV} . \eeqa Because the compactification scale is relatively high
(higher than $150$ TeV), the low-energy effective theory is the
supersymmetric left-right model.

  If the compactification scale $M_C$ is lower than $150$ TeV,
  strongly coupled unification can occur. For example,
if we chose TeV-scale extra-dimension with $M_C=5.0$
  TeV while $M_R=2.0$ TeV, the strongly coupled unification
scale (identify as the cut off
  scale $\Lambda$) is $M_U \sim 30 M_C \sim 150$ TeV.
We can see that $M_C\sim 0.01
  \Lambda$ so that the uncertainties from brane kinetic terms are very small.

In non-SUSY cases, the low-energy left-right model contains one
bi-doublet, one $SU(2)_L$ triplet and one $SU(2)_R$ triplet. The bulk
Higgses contain two ${\bf \overline{10}}$ dimensional representations with
parity assignments $\eta=1$ and $\eta^\pr=\pm1$ \footnote{As noted
previously, we cannot introduce a ${\bf 6}$ representation Higgs only,
because the low energy mass spectrum is not acceptable.}. We obtain the
following beta functions for the gauge couplings of $SU(2)_L$ and
the normalized $U(1)_{B-L}$:
\beqa
 (b_1,b_2,b_3)&=&\(~7,-3,-7\) ~~~~~~~~~~{\rm for}~~ M_Z<E<M_R~, \\
 (b_1,b_2^L,b_2^R,b_3)&=&\(~\f{7}{3},-\f{7}{3},-\f{7}{3},-7 \)
~~{\rm for}~~ M_R<E<M_C ~.
\eeqa
The beta functions of the KK modes are
\beqa
 (b_{B-L,e}, b_{B-L,o})&=&( 1,-13)~, \\
 (b_{2,e}^L, b_{2,o}^L)&=&(-6,-6)~, \\
 (b_{2,e}^R, b_{2,o}^R)&=&(-6,-6)~, \\
(b_{3,e}^R, b_{3,o}^R)&=&(-21/2,0)~. \eeqa The beta functions of
$SU(2)_R$ are the same as those of $SU(2)_L$ due to the left-right
symmetry.

In the non-SUSY case, the power law running with negative beta
functions drive the gauge couplings asymptotically free. We assume
here the unification scale $M_U$ is less than the cut off scale
$\Lambda$. In this case, there are still some allowed parameter
space for the values of $M_C$ and $M_R$ which admit gauge coupling
unification. In fact, $M_C$ is allowed to be as low as $3.0$ TeV
with $M_R=2.2$ TeV (although it is not natural in non-SUSY case to
get such low $M_R$). However, the numerical calculations indicate
that the successful gauge coupling unification requires the
compactification scale $M_C$ to be lower than $8.0$ TeV. For $M_C$
higher than $8.0$ TeV, the $SU(2)_L$ and $U(1)_{B-L}$ gauge
couplings tend to be zero asymptotically without intersection.
Choosing $M_R=3.0$ TeV and $M_C=5.0$ TeV, we obtain \beqa
 \alpha_{B-L}(M_R)=31.6011~, ~~~~
 \alpha_{L}(M_R)=\alpha_{R}(M_R)=31.2399~,
\eeqa
and a unification scale much lower than previously
\beqa
 M_U=5.2473 {\rm ~TeV}~.
\eeqa
In this scenario the relatively low left-right and
compactification scales allow for a unification scale of several
TeV. Although low $M_R$ scenario needs fine-tunning, it is however
possible. Such low-energy unification may have numerous interesting
phenomenological consequences.

The generic phenomenology of our model is similar to that
of any other theories with an extra dimension and thus is not discussed
here. But our model has some additional phenomenological features.
The scenario predicts the existence of doubly
charged gauge bosons at several TeVs which may be within the reach of
the LHC. These heavy gauge bosons have gauge couplings to leptons while
have no couplings to quarks. Since the $(+,-)$ modes vanish on the
$O^\pr$ brane, they can only have derivative couplings to quarks.
But two quark interactions with $A_\mu^X$ are forbidden because of
non-matching quantum numbers. From the mode expansion of the gauge
couplings to leptons, which is similar to that of the Yukawa couplings, we can
see that the doubly charged heavy gauge boson $A^{--}$ can couple to
two charged leptons. It can decay into electron pairs or a pair of
$SU(2)_L$ and $SU(2)_R$ charged gauge bosons $W^-_{1}$ and
$W^{-}_2$. The coupling of the first KK excitations
of the real scalar $A_5^a$ with the leptons can also give couplings
of the charge-two real scalar to charged lepton pairs. We know that
$\phi_3$ is non-vanishing on the $O^\pr$ brane because it has parity
$(-,+)$ under projection. Similar to heavy gauge boson cases, its
couplings to two quarks are forbidden because of its non-matching
quantum numbers. Our model also have $SU(2)_L$ singlets charged
scalars with $B-L=\pm 2$. Such scalars can decay into lepton pairs
like $\nu_{e}\mu$.

\section{$SU(4)_W$ Breaking to $SU(2)_L\tm U(1)_{3R} \tm U(1)_{B-L}$}
\label{sec-4}

  As we demonstrated it is advantageous to break the $SU(4)_W$ to the minimal
left-right model via orbifolding, and the corresponding OGSB chain
for $SU(4)_W$ can be fairly rich. In this section we show that we
can break $SU(4)_W$ to $SU(2)_L\tm U(1)_{3R} \tm U(1)_{B-L}$ which
also leads to interesting phenomenology.

  In this case, our starting point is again the five dimensional $N=1$
supersymmetric $SU(3)_C\tm SU(4)_W$ gauge symmetry.
First, we consider the parity assignments in term of the fundamental
representation of $SU(3)_C \tm SU(4)_W$:
\beqa
P&=&{\rm diag}(+1,+1,+1)\otimes {\rm diag}(+1,+1,+1,-1) ~,~~~\nonumber\\
P^{\pr}&=&{\rm diag}(+1,+1,+1)\otimes {\rm diag}(+1,+1,-1,-1)~.~\,
\eeqa
Boundary conditions break $N=2$ supersymmetry to $N=1$ in four
dimensions. The $SU(3)_C\tm SU(3)_W \tm U(1)_{1}$ gauge symmetry is
preserved at the $O$ brane while it is broken to $SU(3)_C \tm
SU(2)_L \tm SU(2)_R \tm U(1)_{B-L}$ at the $O^\pr$
brane \footnote{In fact, various combinations of $U(1)$ Abelian
groups may remain, since the inner automorphism OGSB will not reduce
the rank of the groups while $Z_n$ orbifolding \cite{hebecker}. For
example, the $U(1)_1$ and the diagonal
components $T^8$ of $SU(3)_W$ can be recombined into
$U(1)_{B-L}$ and $U(1)_{3R}$.}.
The zero modes preserve the $SU(3)_C \tm SU(2)_L \tm U(1)_{3R} \tm
U(1)_{B-L}$ gauge symmetry, which can be seen through the form of
corresponding generators in $SU(4)_W$.

  We introduce two $N=2$ Higgs hypermultiplets in $SU(3)_C\tm SU(4)_W$
symmetric representations in the bulk. These contain the $N=1$
chiral supermultiplets $\Phi^1(1,10)$ and $\Phi^2(1,\overline{10})$
as well as their conjugate chiral fields. The parity assignments for
the Higgs sector read
\beqa
 \eta_{\Phi^i}=1~, ~~~~~~~~~~ \eta'_{\Phi^i}=-1~.
\eeqa
This leads to the following parity assignments for the Higgs hypermultiplets
\beqa
\Phi^i&=&\left(\bea{cccc}(+,-)&(+,-)&(+,+)&(-,+)\\(+,-)&(+,-)&(+,+)&(-,+)\\
(+,+)&(+,+)&(+,-)&(-,-)\\(-,+)&(-,+)&(-,-)&(+,-)\eea
\right)~,~ \nn \\
(\Phi^i)^c&=&\left(\bea{cccc}(-,+)&(-,+)&(-,-)&(+,-)\\(-,+)&(-,+)&(-,-)&(+,-)\\
(-,-)&(-,-)&(-,+)&(+,+)\\(+,-)&(+,-)&(+,+)&(-,+)\eea
\right)~.~\,
\eeqa
The $SU(2)_L$ doublets $H_u$ and $H_d$ arise from
the bulk zero modes of $\Phi^i$, and two $SU(2)_L$ singlets $T_1$
and $T_2$ from that of $(\Phi^i)^c$.

  Fermions can be located at the fix points $O$ or
$O^\pr$. Since at $O$ the gauge symmetry is $SU(3)_C\tm SU(3)_W\tm
U(1)_1$, if we place all the matter on the $O$ brane, we have to
introduce mirror fermions for quarks similarly to the 3-3-1 model.
Thus, the most economical way is to locate all matter at the $O^\pr$
brane (although it is also possible to put leptons on the $O$ brane
while quarks are on $O^\pr$ brane). Since at $O^\pr$ only the
$SU(2)_L\tm SU(2)_R \tm U(1)_{B-L}$ gauge symmetry is preserved,
we can start with a left-right gauge
invariant Lagrangian and then integrate out the heavy modes to get
the $SU(2)_L \tm U(1)_{3R} \tm U(1)_{B-L}$ Lagrangian in four
dimension.

  The matter content at the $O^\pr$ brane can be that of the minimal
left-right model:
\beqa
{\bf (3,2,1)}:Q_L&=&\left(\bea{c}u_L\\d_L\eea\right)~,~~~~~~~~~~~~
{\bf (\bar{3},1,\bar{2})}:Q_L^c=\left(\bea{c}u_L^c\\d_L^c\eea\right)~,\\
 {\bf (1,2,1)}:L_L&=&\left(\bea{c}\nu_L\\e_L\eea\right)~,~~~~~~~~~~~~~
 {\bf (1,1,\bar{2})}:L_L^c=\left(\bea{c}\nu_L^c\\e_L^c\eea\right)~.
\eeqa
In terms of $SU(2)_L\tm SU(2)_R\tm U(1)_{B-L}$
representations the Higgses can be written as:
\beqa
\Phi^1({\bf 10})=\left(\bea{cc}A_{2\tm 2}&\phi_{2\tm 2}\\\phi_{2\tm 2}&B_{2\tm
2}\eea \right)~,
~~~~~~~~\Phi^2({\bf \bar{10}})=\left(\bea{cc}A_{2\tm 2}^\pr&\phi_{2\tm
2}^\pr\\\phi_{2\tm 2}^\pr&B_{2\tm 2}^\pr\eea \right)~.~~~
\eeqa
\beqa
(\Phi^1)^c({\bf \overline{10}})=\left(\bea{cccc}A_1&S_1&D_{11}&D_{12}\\
S_1&A_2&D_{21}&D_{22}\\D_{11}&D_{21}&A_3&T_1\\
D_{12}&D_{22}&T_1&A_4 \eea \right),
~~~(\Phi^2)^c({\bf 10})=\left(\bea{cccc}B_1&S_2&E_{11}&E_{12}\\
S_2&B_2&E_{21}&E_{22}\\E_{11}&E_{21}&B_3&T_2\\
E_{12}&E_{22}&T_2&B_4 \eea \right).
\eeqa
The parity of the brane fields are determined by the requirement that all
the gauge invariant operators on the $O^\pr$ brane must transform covariantly
under $P$ parity which is the consequence of the identification of the
$y={\pi R}/2$ and $y=-{\pi R}/2$ branes. From the kinetic terms and parity
assignments follows the parity of the matter content on the $O^\pr$
brane:
\beqa
P:~Q_L&=&\pm(+,+)~,~~~~~~~~~~~~~~~P:~ L_L=\pm(+,+)~, \nm\\
P: ~Q_L^c&=&\pm(+,-)~,~~~~~~~~~~~~~~~P:~ L_L^c=\pm(+,-)~.
\eeqa
From the parity properties of gauge invariant operators on the
$O^\pr$ brane (which we do not list here) the Yukawa couplings of
the bulk Higgses to the brane fermions can be obtained
\small
\beqa
&& {\cal L}_5=\int d^2\theta \f{\sqrt{\pi R}}{2} \nn \\
&&~\times\left\{\f{}{}\right.\left[\delta(y-\f{\pi R}{2}) \pm
\delta(y+\f{\pi R}{2})\right]
 \sum\limits_{ij} \( Y^1_{ij} \epsilon^{ab}
   (Q_L)_{a}^{i}\(Q_L^c\)_{c}^j (\phi)^{c}_b
 +Y^2_{ij}\epsilon^{bc}
   (Q_L)_{a}^{i}(Q_L^c)_{c}^j(\phi^{\pr})^a_b \f{}{} \)\nonumber\\
&&~+\f{1}{2}\left[\delta(y-\f{\pi R}{2}) \pm \delta(y+\f{\pi R}{2})\right]
\sum\limits_{ij} \( Y^3_{ij} \epsilon^{ab}
   (L_L)_{a}^{i}(L_L^c)_{c}^j(\phi)^{c}_b
+Y^4_{ij}\epsilon^{bc}
   (L_L)_{a}^{i}(L_L^c)_{c}^j(\phi^{\pr})^a_b \)\nonumber\\
&&~+\f{1}{2}\left[\delta(y-\f{\pi R}{2}) + \delta(y+\f{\pi
R}{2})\right] \sum\limits_{ij} \( Y^5_{ij} \epsilon^{ab}
   (L_L)_{a}^{i}(L_L)_{b}^jT_1
+Y^6_{ij}\epsilon^{ab}
   (L_L^c)_{a}^{i}(L_L^{c})_b^jT_2\)\left. \frac{}{}\right\} ~~~~~
\eeqa
\normalsize
Here the $\pm$ signs correspond to the relative parity
assignment (identical or inverse) in front of $Q_L$ and $Q_L^c$
($L_L$ and $L_L^c$), respectively. After expanding $\phi$ and
$\phi^c$ in their KK modes, we can see that amongst the zero modes
only two $SU(2)_L$ doublets remain which are identified as $H_u$ and
$H_d$ of the supersymmetric standard model.The electric charged
field $T_1$($T_2$) can couple to two leptons as $\nu_L e_L$($\nu_R
e_R$) etc.
The $U(1)_{B-L}$ number for quarks and leptons can be determined
by anomaly cancellation requirements for $[SU(2)_L]^2U(1)_{B-L}$,
$[SU(2)_R]^2U(1)_{B-L}$, $[U(1)_{B-L}]^3$, as well as $[{\rm
Gravity}]^2U(1)_{B-L}$, and $[SU(3)_C]^2U(1)_{B-L}$, etc.
The normalization of the Higgs sector can be determined by the
requirement that the Yukawa couplings should be invariant under
$U(1)_{B-L}$. The charge quantization conditions, in
terms of the $SU(4)_W$ fundamental representation, are
\beqa
  Q_1+Q_3=Q_2+Q_3=0, ~~~~~~ Q_3+Q_4=2b ~,
\eeqa
where $b$ is the $U(1)_{B-L}$ number for leptons. The fields
$T_1$ and $T_2$ are necessary to determine the $U(1)_{B-L}$
quantization conditions because they give the second equation in the
previous formula. From the first equation and traceless condition follows
that the $U(1)_{B-L}$ generator is proportional to the $SU(4)_W$
generator
\begin{eqnarray}
 T_{B-L}~=~ {\rm diag}(-a,-a,~a,~a) ~.
\end{eqnarray}
From the second equation we obtain that $a=b=1$.  (Here we rely on
the phenomenological requirement that the relative normalization of
the $U(1)_{B-L}$ charge between the Higgs and lepton sectors was
chosen to be $b=1$). From these quantization conditions, we obtain
 $2g_{B-L}^2=g_4^2$. Since the $U(1)_{3R}$ gauge group can be realized as the
diagonal subgroup of $SU(2)_R$, its normalization condition is set
by $SU(2)_R$, which leads to the relation
$g_{3R}^2=g_4^2$.
From the charge assignments we obtain
\beqa
 Q=T_{3L}+\f{Y_{3R}}{2}+\f{Y_{B-L}}{2}~.
\eeqa
The tree level weak mixing angle is again $\sin^2\theta_W=0.25$.

  The quantization conditions imply the parity and quantum
numbers for all the bulk fields
\small
\beqa
&&V({\bf 15})={\bf 3_{(0,0)}^{(+,+)}\oplus 1_{(0,0)}^{(+,+)}\oplus
1_{(0,0)}^{(+,+)}\oplus 2_{(-1,-2)}^{(+,-)}\oplus{\bar 2}_{(1,2)}^{(+,-)}
\oplus 2_{(1,-2)}^{(-,-)}\oplus{\bar
2}_{(-1,2)}^{(-,-)}\oplus 1_{(-2,0)}^{(-,+)}\oplus 1_{(2,0)}^{(-,+)}}\nonumber\\
&&\Sigma({\bf 15})={\bf 3_{(0,0)}^{(-,-)}\oplus
1_{(0,0)}^{(-,-)}\oplus 1_{(0,0)}^{(-,-)}\oplus
2_{(-1,-2)}^{(-,+)}\oplus{\bar 2}_{(1,2)}^{(-,+)}\oplus
2_{(1,-2)}^{(+,+)}\oplus{\bar 2}_{(-1,2)}^{(+,+)}\oplus
1_{(-2,0)}^{(+,-)}\oplus 1_{(2,0)}^{(+,-)}}\nonumber\\
&&\Phi({\bf 10})={\bf 3_{(0,-2)}^{(+,-)}\oplus
2_{(1,0)}^{(+,+)}\oplus 2_{(-1,0)}^{(-,+)}\oplus
1_{(2,2)}^{(+,-)}\oplus 1_{(-2,2)}^{(+,-)}\oplus 1_{(0,2)}^{(-,-)}},\nonumber\\
&& \Phi^c({\bf \overline{10}})={\bf \bar{3}_{(0,2)}^{(-,+)}\oplus
\bar{2}_{(-1,0)}^{(-,-)}\oplus \bar{2}_{(1,0)}^{(+,-)}\oplus
1_{(-2,-2)}^{(-,+)}\oplus 1_{(2,-2)}^{(-,+)}\oplus
1_{(0,-2)}^{(+,+)}},\nm\\
&& \Phi({\bf \overline{10}})={\bf \bar{3}_{(0,2)}^{(+,-)}\oplus
\bar{2}_{(-1,0)}^{(+,+)}\oplus \bar{2}_{(1,0)}^{(-,+)}\oplus
1_{(-2,-2)}^{(+,-)}\oplus 1_{(2,-2)}^{(+,-)}\oplus
1_{(0,-2)}^{(-,-)}},\nonumber\\
&&\Phi^c({\bf {10}})={\bf 3_{(0,-2)}^{(-,+)}\oplus
2_{(1,0)}^{(-,-)}\oplus 2_{(-1,0)}^{(+,-)}\oplus
1_{(2,2)}^{(-,+)}\oplus 1_{(-2,2)}^{(-,+)}\oplus 1_{(0,2)}^{(+,+)}}.
\eeqa
\normalsize
Subscripts denote $U(1)_{3R}$ and $U(1)_{B-L}$ quantum
numbers, respectively. We can see that there are zero mode
components in $\Sigma({\bf 15})$ decompositions. Such zero modes can
act as Higgs doublets in the MSSM, if we adopt the gauge-Higgs
unification scheme. However such Higgs fields cannot couple to
matter fields because of un-matching quantum numbers.

  The $SU(2)_L\tm U(1)_{3R}\tm U(1)_{B-L}$ gauge symmetry can be
broken to the SM one (in SUSY cases) via the bulk Higgs
fields $H^1{\bf (1,4)}$ and $H^2{\bf (1,\bar{4})}$
(here $SU(3)_C\tm SU(4)_W$ representations are shown).
Parity can be assigned to these Higgses as
\beqa
 \eta_{H^i}=-1~,~~~~~~~~~~~\eta'_{H^i}=-1~ ~~~~~(i=1,2)~.
\eeqa

From the decomposition of $SU(4)_W$ in terms of $SU(2)_L\tm
U(1)_{3R}\tm U(1)_{B-L}$
\small
\beqa (H^1)({\bf 4})={\bf
2}_{(0,-1)}^{(-,-)}\oplus{\bf 1}_{(1,1)}^{(-,+)}\oplus{\bf
1}_{(-1,1)}^{(+,+)}~,~~~~~(H^1)^c({\bf \bar{4}})={\bf
\bar{2}}_{(0,1)}^{(+,+)}\oplus{\bf 1}_{(-1,-1)}^{(+,-)}\oplus{\bf
1}_{(1,-1)}^{(-,-)}~,\nonumber\\
(H^2)({\bf \bar{4}})={\bf \bar{2}}_{(0,1)}^{(-,-)}\oplus{\bf
1}_{(-1,-1)}^{(-,+)}\oplus{\bf 1}_{(1,-1)}^{(+,+)}~,~~~(H^2)^c({\bf
4})={\bf 2}_{(0,-1)}^{(+,+)}\oplus{\bf 1}_{(1,1)}^{(+,-)}\oplus{\bf
1}_{(-1,1)}^{(-,-)}~,~~
\eeqa
\normalsize
follows that the zero modes of
$H^i~(i=1,2)$ contain two $SU(2)_L$ singlets $U^1_{\bf(-1,1)}$ and
$U^2_{\bf(1,-1)}$ (subscripts denote $U(1)_{3R}\tm U(1)_{B-L}$
quantum numbers) which are electrically neutral and cannot couple to
matter directly. The zero modes for $(H^i)^c$ contain two Higgs
doublets ${\bf \bar{2}}_{(0,1)}^{(+,+)}$ and ${\bf
2}_{(0,-1)}^{(+,+)}$ which can not couple to matter either because
of non-matching quantum numbers. After $U^1$ and $U^2$ acquire VEVs,
the
remaining gauge symmetry is broken to the SM gauge group
\footnote{It is also possible to break the remaining gauge group to
the SM via localized Higgs fields $A({\bf 1,2,-1})$ (in terms of
$SU(2)_L\tm SU(2)_R\tm U(1)_{B-L}$ quantum number) on the $O^\pr$
brane. Such localized brane Higgs fields can break the gauge group
$SU(2)_R\tm U(1)_{B-L}$ on the $O^\pr$ brane to $U(1)_Y$, which
corresponds to breaking the bulk $U(1)_{B-L}\tm U(1)_{3R}$ to
$U(1)_Y$. The other possibility is to introduce two $\Delta({\bf
\overline{10}})$ representations for $SU(4)_W$ with parity
assignment
 $\eta_{\Delta^i}=\eta_{\Delta^i}^\pr=1$.
The VeV of the neutral component ${\bf 1_{(2,-2)}^{(+,+)}}$ will
break $U(1)_{3R}$ as well as give Majorana neutrino masses for right
handed neutrino.}.
Note that $T_1$ and $T_2$ cannot be used to break
this gauge symmetry because they have electric charges.

The beta functions of the gauge couplings $U(1)_Y,~SU(2)_L,~SU(3)_C$
read
\beqa
(b_1,b_2,b_3)&=&\(\f{25}{3},-\f{7}{3},-7\)~~~~~~{\rm for}~~~M_Z<E<M_S~, \\
(b_1,b_2,b_3)&=&\(~15,~~3,-3 \)~~~~~~~~{\rm for}~~~M_S<E<M_Z^{\pr}~.
\eeqa
In the SUSY and SUSY decoupling limits, there are six Higgs
doublets.
For the $\sqrt{2}U(1)_{B-L}$, $U(1)_{3R}/2$, $SU(2)_L$, and
$SU(3)_C$ gauge couplings the beta functions are
\beqa
(b_1^{B-L},b_1^{3R},b_2^L,b_3)=\(\f{23}{4},\f{17}{2},~3,-3
\)~~{\rm for}~~~M_Z^\pr<E<M_C ~.
\eeqa
The beta functions corresponding to the even and odd KK modes at
one loop are
\beqa
 (b_{B-L,e}, b_{B-L,o})&=&(-\f{1}{2},\f{13}{2}) ~, \\
 (b_{3R,e}, b_{3R,o}^L)&=&(~1,~5) ~, \\
 (b_{2,e}^R, b_{2,o}^R)&=&(-2,~8) ~,\\
 (b_{3,e}^R, b_{3,o}^R)&=&(-6,~0) ~.
\eeqa

  Just as in the previous case, the $SU(4)_W$ preserved $b_o+b_e$
is constant for the three gauge couplings. Thus, the relative RGE
running between the three gauge couplings are logarithmic. As we do
not know the $g_{B-L}$ or $g_{3R}$ gauge couplings at the $M_Z^\pr$
scale (or the relations between the two gauge couplings), we must
invoke further assumptions related to them to predict the
unification scales. SUSY breaking can again be achieved by the
Scherk-Schwarz mechanism through boundary conditions. The tree-level
gaugino and Higgsino masses acquired this way will induce loop-level
squark and slepton masses.

  The phenomenology of this symmetry breaking chain shares many common
features with that of the previous cases. For example, there are
charge two heavy gauge bosons and two $SU(2)_L$ charged gauge
singlets scalars which can only derivatively couple to charged
lepton pairs. At energies well below the $M_C$ scale, the low
energy effective theory reduces to supersymmetric $SU(2)_L\tm
U(1)_{3R}\tm U(1)_{B-L}$. This $U(1)$ extension of the SM has
been widely studied. The special feature of this scenario is the
existence of Higgs doublets which have no tree level couplings to
SM fermions even when the low energy $SU(2)_L\tm U(1)_Y$
quantum number allow such couplings. The electrically neutral $SU(2)_L$ singlet
Higgses, $U^1$ and $U^2$, which break the remaining group to the SM,
can be viable dark matter candidates.
\vspace*{0.5cm}

\noindent {\bf Alternative Models:~~}
  We can locate the SM quarks and right-handed charged leptons on the
$O'$ brane while placing the SM lepton doublets and
right-handed neutrinos in the bulk. We can introduce mirror leptons
$X_L$, $X_L^c$, $\bar{X}$, $\bar{X}^c$, $Y_L$, and $Y_L^c$ to fill the bulk
hypermultiplets $F_{i}~(i=1,2)$ in the
$\mathbf{(1,4)}$ representation under $SU(3)_C\times SU(4)_W$:
\beqa
F_1=(L_L~X_L)~,~~~~~~F_2=(\bar{X},(L^c_L)^\pr)~, \nm\\
F_1^c=(Y_L~X_L^c)~,~~~~~~F_2^c=(\bar{X}^c,Y_L^c) ~. ~~~
\eeqa
Here $(L^c_L)^\pr$ denotes $(E_L^c~-\nu^c_L)$, with $E_L^c$ being a
charged mirror lepton. Then, we can assign parities as
\begin{eqnarray}
 \eta_{F_{1}}=1~,~~~\eta'_{F_{1}}=1~,~\,~~\eta_{F_{2}}=-1~,~~~\eta'_{F_{2}}=-1 ~.
\end{eqnarray}
The left-handed leptons and neutrinos $L_L$ arise from $F_1$,
and the right-handed neutrinos from $F_2$. Note that we cannot fit
right-handed leptons in $F^c_1$ because that does not yield the correct
quantum numbers. Mirror fermions associated with each SM
leptons, except with the right-handed charged leptons, will survive the
projection.

  As previously, we can locate the SM quarks and right-handed
neutrinos on the $O'$ brane while having the SM lepton
doublets and right-handed charged leptons in the bulk. The parity
assignments read
\beqa
P&=&{\rm diag}(+1,+1,+1)\otimes {\rm diag}(+1,+1,-1,+1)~, \nonumber\\
P^{\pr}&=&{\rm diag}(+1,+1,+1)\otimes {\rm diag}(+1,+1,-1,-1)~.
\eeqa
Similarly to our previous case, we obtain mirror fermions
associated with each SM leptons from zero modes, except for
right-handed neutrinos.

\section{Conclusions}
\label{sec-5}
  In this paper, we propose a low scale $SU(4)_W$ unification model
which has two symmetry breaking chains. In the first chain $SU(4)_W$
is broken into the $SU(2)_L\times SU(2)_R\times U(1)_{B-L}$ minimal
left-right model through $S^1/(Z_2\tm Z_2)$ orbifolding. Leptons are
fitted into $SU(4)_W$ multiplets and located on a symmetry
preserving $O$ brane, while quarks are placed on $O^\pr$ brane where
the symmetry is broken. This approach predicts $\sin^2\theta_W=0.25$
for the weak mixing angle at tree level and leads to a rather low
weakly coupled unification scale of order $10^2$ TeV with
supersymmetry, or as low as several TeV in the non-supersymmetric
case. If we introduce mirror fermions and put quarks in the bulk,
the model gives a large weak mixing angle $\sin^2\theta_W=0.45$
which will lead to high-energy unification. The other symmetry
breaking chain with the low-energy gauge group $SU(2)_L\tm U(1)_{3R}
\tm U(1)_{B-L}$ after OGSB can also give rise to a weak mixing angle
$\sin^2\theta_W=0.25$ at tree level. In this scenario, leptons and
quarks are placed on the $O^\pr$ brane (with broken symmetry) and
the quantization conditions are determined by anomaly cancelation
requirements. These low-scale unification theories have interesting
phenomenological consequences.

One may worry if there are cosmological difficulties associated with
this scenario such as the monopole problems etc. In fact there are
no monopole problems in our scenario because we break the gauge
symmetry via orbifolding. In general there are monopole problems if a
gauge symmetry is broken to a subgroup containing $U(1)$ via Higgs
mechanism with the unification scale lower than the inflation scale
and at the same time higher
than TeV scale \cite{hitoshi}. It is not a problem in OGSB scenario
 because the gauge symmetry is broken via boundary conditions with
 the symmetry broken explicitly in the orbifold fix points. So our
 scenario is not bothered by the cosmological monopole problems.

\section*{Acknowledgements}
We acknowledge Jing Shu and Hitoshi Murayama for comments. This
work was supported by the Australian Research Council under project
DP0877916, by the National Natural Science Foundation of China under
grant Nos. 10821504, 10725526 and 10635030, and by the
Cambridge-Mitchell Collaboration in Theoretical Cosmology.

\end{document}